\documentclass[pra,aps,twocolumn]{revtex4}
\usepackage{epsfig,latexsym,amssymb,amsmath,amsbsy,graphics,graphicx}

\def \beq{\begin{equation}}
\def \eeq{\end{equation}}
\def \beqarr{\begin{eqnarray}}
\def \eeqarr{\end{eqnarray}}

\begin{document}

\title{Fermi surfaces and Luttinger's theorem in paired fermion systems}

\author{Subir Sachdev}

\affiliation{Department of Physics, Harvard University, Cambridge,
Massachusetts 02138, USA}

\author{Kun Yang}
\thanks{On leave from Department of Physics, Florida State University,
Tallahassee, FL 32306.}

\affiliation{Department of Physics, Harvard University, Cambridge,
Massachusetts 02138, USA}

\date{\today}

\begin{abstract}
We discuss ground state properties of a mixture of two fermion
species which can bind to form a molecular boson. When the densities
of the fermions are unbalanced, one or more Fermi surfaces can
appear: we describe the constraints placed by Luttinger's theorem on
the volumes enclosed by these surfaces in such Bose-Fermi mixtures.
We also discuss the nature of the quantum phase transitions
involving changes in the number of Fermi surfaces.
\end{abstract}

\maketitle

\section{Introduction}
\label{sec:intro}

Recent experiments \cite{martin,randy} on trapped ultracold atoms
have refocused theoretical attention on pairing between distinct
fermion species, in situations in which the densities of the two
species are unequal. In these experiments unequal numbers of two
hyperfine states of fermionic $^6$Li atoms were mixed and scanned
across a Feshbach resonance. On one side of the resonance the
fermions bind to form a bosonic molecule which can condense into a
Bose-Einstein condensate (BEC), while on the other side they
experience a weak attractive interaction which leads to formation of
Cooper pairs in a Bardeen-Cooper-Schrieffer (BCS) state. When the
densities of the hyperfine species are unbalanced, some fermions are
unpaired, and this raises the possibility of co-existence of a
bosonic condensate and one or more Fermi surfaces in the ground
state.

This paper will describe the constraints that Luttinger's theorem
places on the volumes enclosed by Fermi surfaces under such
conditions. Such constraints are distinct in different phases, and
we will also describe the quantum phase transitions across which the
statement of Luttinger's theorem changes. We note that these Fermi
surface volume constraints are exact, and apply even in strongly
interacting regimes where the fermions fluctuate into bosonic
molecular states at short scales. In such situations short-range
spectroscopic observables may indicate that fermions exist in
molecules, but the true ground state will nevertheless have the
undiminished Fermi surface volume(s), albeit with a reduced
quasiparticle residue at the Fermi surface. Our results indicate
that the volumes of the Fermi surfaces are sensitive to the presence
or absence of a Bose condensate of the molecules; thus they can also
be used to probe superfluidity or pair coherence. In principle the
Fermi surface volumes can be measured from the momentum
distributions of the atoms. Such a measurement was recently
performed in a gas of $^{40}$K across a Feshbach
resonance\cite{regal}. In this experiment the effect of the trap on
the momentum distribution appears to be quite strong, such that the
discontinuity in momentum distribution gets wiped out even for
non-interacting fermions. We hope that by manipulating the form of
the trap potential, its effect can be minimized so that
discontinuities in momentum distribution associated with Fermi
surfaces can be detected in future experiments; this would probably
require a trap potential that is flat inside the trap and rises very
fast near the boundary.

We note that Luttinger's theorem for Bose-Fermi mixtures has also
been investigated in recent work \cite{powell,coleman} in which a
boson and a fermion bind to form a fermionic molecular state. Here
we will show that these results can be straightforwardly extended to
the case of interest to the recent atomic experiments: two fermions
binding to form a bosonic molecular state. We also use
non-perturbative arguments similar to those of Yamanaka, Oshikawa
and Affleck \cite{yoa} to establish analogous results in one
dimension.

\section{Hamiltonian and phase diagram}
\label{sec:ham}

Our results are rather general, and apply to a wide class of
Hamiltonians of the form
\begin{eqnarray}
 H &=& H_f+H_b+\sum_{{\bf k}, {\bf k}'}(V_{{\bf k}, {\bf k}'}
f_{\uparrow,{\bf k}}^\dagger f_{\downarrow, {\bf k}'}^\dagger
b_{-{\bf k}-{\bf k}'}+ h.c.)  \nonumber \\
&-& \sum_{{\bf k}} \left( (\mu + h) f_{\uparrow,{\bf k}}^\dagger
f_{\uparrow, {\bf k}} + (\mu - h) f_{\downarrow,{\bf k}}^\dagger
f_{\downarrow, {\bf k}} \right. \nonumber \\
&~&~~~~~~~~~~~~ \left. +(2\mu - \nu) b_{{\bf k}}^{\dagger} b_{{\bf
k}} \right) , \label{model2}
\end{eqnarray}
where $f_{\uparrow,\downarrow}$ are annihilation operators of the
two fermion species, $b$ is the annihilation operator of the bosonic
molecule, $H_f$ and $H_b$ involve fermionic and bosonic operators
only (which contain their kinetic energies and may also include
additional interactions). The fermion species are at chemical
potentials $\mu_{f \uparrow} = \mu + h$ and $\mu_{f \downarrow} =
\mu - h$, while the boson is at chemical potential $\mu_b = 2\mu -
\nu$. Here $h$ is the `field' which unbalances the fermion
densities, and $\nu$ is the `detuning' across the Feshbach resonance
which scans between the BCS and BEC limits.

\begin{figure}
\includegraphics[width=2.5in]{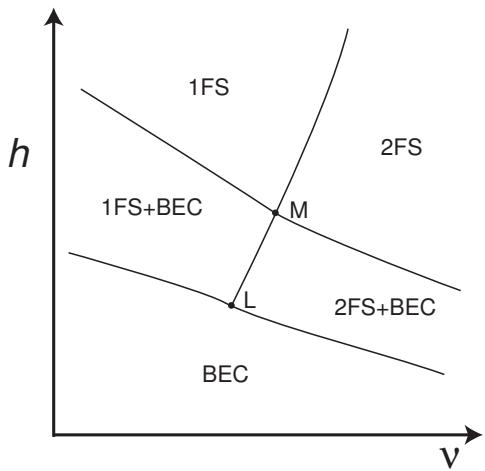}
\caption{Schematic zero temperature phase diagram as a function of
the `field' $h$ and the detuning $\nu$. The phases labeled BEC have
$\langle b \rangle \neq 0$, and include regions where the pairing is
more properly considered to be of the BCS type. States labeled $n$FS
have $n$ Fermi surfaces. Only homogeneous, translationally invariant
states are shown, and possible instabilities to phase separation and
modulated FFLO states have been omitted. Two Luttinger theorems, in
Eq.~(\ref{lutt2fs}), apply in the phases without the BEC. The phases
with the BEC have only one Luttinger theorem, specified in
Eq.~(\ref{lutt1}).} \label{phase}
\end{figure}

It is useful to frame our discussion using a ground state phase
diagram as a function of $\nu$ and $h$. Such a phase diagram has
been studied in a number of recent works
\cite{pao,son,sheehy0,yang,ys,stoof}, and it is not our purpose here
to critique these results. Rather we find it useful to work with the
{\em schematic} phase diagram shown in Fig.~\ref{phase}, which omits
instabilities towards phase separation and broken translational
invariance.
As we will discuss later, our results are easily extended to
modulated phases like the Fulde-Ferrell-Larkin-Ovchinnikov (FFLO)
state \cite{ff,lo}.

It is easiest to first understand the phases at small and large $h$
in Fig~\ref{phase}. At small $h$ we have the conventional paired
fermion state with equal fermion densities, $N_\uparrow =
N_\downarrow$, where $N_\uparrow = \sum_{\bf k}f^\dagger_{\uparrow,
{\bf k}}f_{\uparrow, {\bf k}}$ and similarly for $N_\downarrow$;
this interpolates from the BEC to the BCS limits with increasing
$\nu$. At very large $h$, we have conventional Fermi liquid states
with no BEC, $\langle b \rangle = 0$; these can have a single Fermi
surface of $f_\uparrow$ alone with $N_\downarrow =0$, or two Fermi
surfaces with one each for $f_\uparrow$ and $f_\downarrow$, and both
$N_\uparrow$, $N_\downarrow$ non-zero.

At intermediate $h$, we have the possibility of phases with
co-existence of the BEC and Fermi surface(s). These can be
understood most simply by adding a non-zero $\langle b \rangle$ to
the effective Hamiltonian for the large $h$ Fermi liquid states. The
2FS+BEC phase is known as the Sarma \cite{sarma} or ``breached
pair'' \cite{liu} state. This intermediate $h$ region is most likely
to be susceptible to further instabilities towards phase separation
\cite{bedaque} and modulated phases (which we will note later) in
the weak coupling (large positive $\nu$) regime. There is numerical
evidence suggesting it may become stable in the intermediate and
strong coupling regimes \cite{carlson}.

\section{Luttinger's Theorem}
\label{sec:luttinger}

For the decoupled Hamiltonian $H_f + H_b$, the numbers of the two
fermion species are separately conserved, and the original version
of the Luttinger's theorem states that there are two Fermi surfaces
that correspond to the two fermion species, whose volumes are
determined by the numbers of each fermion species. Here we extend
the Luttinger's theorem to include the full $H$, which mixes
fermions with bosons and {\em break} the conservation of the fermion
number for individual fermion species. We find that the two Fermi
surfaces remain to have separately conserved volumes as long as
there is no Bose condensate (or no broken U(1) symmetry). In the
presence of a Bose condensate, the volumes of individual Fermi
surfaces are no longer conserved, but we will now show that the {\em
difference} of the volumes of the two Fermi surfaces remains
conserved, and this difference is determined by the density or
number difference in $\uparrow$ and $\downarrow$ fermions (which
commutes with $H$):
\begin{equation}
\Delta N=N_\uparrow-N_\downarrow=\sum_{\bf k}(f^\dagger_{\uparrow,
{\bf k}}f_{\uparrow, {\bf k}}-f^\dagger_{\downarrow, {\bf
k}}f_{\downarrow, {\bf k}}).
\end{equation}
As we noted earlier, closely related results were obtained on a
different model of Bose-Fermi mixture \cite{powell}.

We introduce the $2\times 2$ Green's function matrix for fermions in
the usual manner to allow for pairing and appearance of anomalous
Green's function:
\begin{equation}
\hat{G}_F({\bf x}, t)=-i\langle T[f_\uparrow({\bf x}, t),
f^\dagger_\downarrow({\bf x}, t)]^T [f^\dagger_\uparrow(0,0),
f_\downarrow(0,0)]\rangle.
\end{equation}
$\Delta N$ is related to the Green's function matrix in the
following manner:
\begin{eqnarray}
&\sum_{\bf k}&(f_{\uparrow{\bf k}}^\dagger f_{\uparrow {\bf k}}+
f_{\downarrow{\bf k}}f^\dagger_{\downarrow{\bf k}})=N_0+\Delta N \nonumber\\
&=&{-iA\over (2\pi)^{d+1}}\int{d\omega d^d{\bf
k}}e^{i\omega0^+}tr\hat{G}_F(\omega, {\bf k}), \label{number}
\end{eqnarray}
where $N_0=\sum_{\bf k} 1$ is the total number of single particle
states within a high momentum cutoff or a band within which the
model is defined, $A$ is the system volume, and $d$ is
dimensionality. Using the Dyson's equation for fermion self-energy
(including anomalous self-energy that is off-diagonal)
\begin{equation}
\hat{\Sigma}_F(\omega, {\bf k})=\omega-\hat{\xi}_{\bf
k}-\hat{G}_F^{-1}(\omega, {\bf k}),
\end{equation}
where $\hat{\xi}_{\bf k}$ is the (diagonal) kinetic energy matrix
for $\uparrow$ fermion and $\downarrow$ hole (each measured from the
chemical potential), one may rewrite Eq. (\ref{number}) as
\begin{eqnarray}
N_0+\Delta N&=&{iA\over (2\pi)^{d+1}}\int{d\omega d^d{\bf
k}}e^{i\omega0^+}tr\left[{\partial\over \partial
\omega}\log\hat{G}_F(\omega, {\bf k})\right.\nonumber\\
&-& \left . \hat{G}_F(\omega, {\bf k}){\partial\over \partial
\omega}\hat{\Sigma}_F(\omega, {\bf k})\right]. \label{number1}
\end{eqnarray}

The proof of Luttinger's theorem relies on the existence of a
Luttinger-Ward (LW) functional $\Phi_{LW}[G']$, whose functional
derivative gives the exact self-energy when evaluated at the exact
Green's function \cite{lw}:
\begin{equation}
{\delta\Phi_{LW}[G']\over \delta G'}|_{G'=G}=\Sigma.
\label{selfenergy}
\end{equation}
Recently it has been shown \cite{potthoff} the LW functional can be
constructed non-perturbatively under very general situations,
including cases with spontaneously broken symmetry. This is
particularly important in the latter cases, as the broken symmetry
states are {\em not} smoothly connected to non-interacting systems,
and the Green's functions may contain ``anomalous" terms. In the
present case the LW functional is a functional of both fermion and
boson Green's functions, and the conservation of $\Delta N$
guarantees that
\begin{equation}
\Phi_{LW}[\hat{G}'_F(\omega, {\bf k}), G'_B(\omega, {\bf
k})]=\Phi_{LW}[\hat{G}'_F(\omega+\alpha, {\bf k}), G'_B(\omega, {\bf
k})];
\end{equation}
this is a consequence of the invariance of
$\Phi_{LW}[\hat{G}'_F(\omega, {\bf k}), G'_B(\omega, {\bf k})]$
under gauge transformation $f_\uparrow\rightarrow f_\uparrow
e^{i\delta}; f_\downarrow\rightarrow f_\downarrow e^{-i\delta}$
\cite{potthoff}. It leads to
\begin{equation}
{\partial\Phi_{LW}[\hat{G}_F', G_B']\over
\partial\alpha}=\int{d\omega d^d{\bf
k}}tr\left[{\delta\Phi_{LW}\over
\delta\hat{G}'_F}{\partial\hat{G}'_F\over \partial\omega}\right]=0.
\label{gauge}
\end{equation}
Now use the exact Green's functions as the argument of $\Phi_{LW}$
in the equation above, and generalize Eq. (\ref{selfenergy}) to the
matrix form:
\begin{equation}
{\delta\Phi_{LW}[\hat{G}'_F, G'_B]\over
\delta\hat{G}'_F}|_{\hat{G}'=\hat{G}}=\hat{\Sigma}_F, \label{sigma}
\end{equation}
one can show the last term in Eq. (\ref{number1}) vanishes by
combining Eqs. (\ref{sigma}) and (\ref{gauge}). The manipulation of
the remaining term (that involve $\hat{G}_F$ only) is
straightforward and standard \cite{lw,luttinger}; one may, for
example, diagonalize $\hat{G}_F$ to perform the trace. This leads to
\begin{equation}
N_0+\Delta N={A\over (2\pi)^{d}}\int{d^d{\bf
k}}[\Theta(-e_\uparrow({\bf k}))+\Theta(-e_\downarrow({\bf k}))],
\label{lutt}
\end{equation}
where $e_{\uparrow,\downarrow}({\bf k})$ are the eigenvalues of
$-\hat{G}^{-1}(\omega=0, {\bf k})$. In the absence of interactions
$e_{\uparrow,\downarrow}({\bf k})$ are simply the single particle
excitation energies, and the ${\bf k}$'s at which they vanish define
the Fermi surfaces. In the presence of interactions there is a
self-energy contribution to $e_{\uparrow,\downarrow}({\bf k})$,
which in general contains an imaginary part. However for the
quasiparticles to be well-defined in the low-energy limit, the
imaginary part of the self-energy must vanish for $\omega=0$, so
that the quasiparticles are long-lived. As a consequence
$e_{\uparrow,\downarrow}({\bf k})$ are real, and are the
(interaction-renormalized) quasiparticle excitation energies, and
the ${\bf k}$'s at which they vanish define the location of the
Fermi surfaces {\em in the presence of interactions}. Thus the two
terms on the right hand side of Eq. (\ref{lutt}) are the volumes of
the two Fermi seas for $\uparrow$ particle-like quasiparticles and
$\downarrow$ {\em hole}-like quasiparticles; this is because
$e_{\uparrow}({\bf k})$ is an increasing function of $|{\bf k}|$,
while $e_{\downarrow}({\bf k})$ is a decreasing one. To express
results in terms of particle-like Fermi seas, we rewrite Eq.
(\ref{lutt}) as
\begin{eqnarray}
\Delta N&=&-N_0+{A\over (2\pi)^{d}}\int{d^d{\bf
k}}[\Theta(-e_\uparrow({\bf k}))+1-\Theta(e_\downarrow({\bf
k}))]\nonumber\\
&=&{A\over (2\pi)^{d}}\int{d^d{\bf k}}[\Theta(-e_\uparrow({\bf
k}))-\Theta(e_\downarrow({\bf k}))]\nonumber\\
&=&{A\over (2\pi)^{d}}(\Omega_\uparrow-\Omega_{\downarrow}),
\label{lutt1}
\end{eqnarray}
where we used the fact that $N_0={A\over (2\pi)^{d}}\int{d^d{\bf
k}}$. We have thus related the fermion number difference $\Delta N$
to the Fermi surface volumes $\Omega_\uparrow, \Omega_{\downarrow}$
in a form analogous to the Luttinger's theorem \cite{luttinger}. The
statement in Eq.~(\ref{lutt1}) applies in {\em all\/} the phases of
Fig.~\ref{phase}. The BEC phase has no Fermi surface, and so in this
phase we must have $N_\uparrow = N_\downarrow$.

In the absence of Bose condensation, $\hat{\Sigma}$ (and therefore
$\hat{G}_F$) is diagonal in the original basis of
$f_{\uparrow,\downarrow}$; in this case one can interpret the two
Fermi surfaces as those for the $\uparrow,\downarrow$ fermions
respectively. By using the fact that bosonic excitations are gapped
in the absence of a Bose condensate \cite{powell}, one can further
prove that their volumes are fixed {\em individually}, so that
\begin{equation}
N_\uparrow + N_B = \frac{A}{(2 \pi)^d} \Omega_\uparrow~~;~~
N_\downarrow +N_B = \frac{A}{(2 \pi)^d} \Omega_\downarrow ,
\label{lutt2fs}
\end{equation}
where $N_B=\sum_{\bf k}b^\dagger_{\bf k}b_{\bf k}$. The result
(\ref{lutt2fs}) applies in all phases of Fig.~\ref{phase} without a
BEC. Among these, the 1FS phase only has a Fermi surface for the
$\uparrow$ fermions, and so Eq.~(\ref{lutt2fs}) implies that in this
phase we must have $N_\downarrow = 0$ and $N_B = 0$. Note that
$N_{\uparrow,\downarrow}$ and $N_B$ are not individually conserved,
but the combinations on the left-hand-sides of Eq.~(\ref{lutt2fs})
are conserved in the absence of a Bose condensate. The proof of
Eq.~(\ref{lutt2fs}) parallels the analogous one in
Ref.~\onlinecite{powell}.

In our discussion so far we have assumed the system to be uniform.
Our results, however, can be generalized to cases with spontaneously
broken translational symmetry. An important example of this case is
the FFLO superfluid state \cite{ff,lo}, in which the Bose condensate
has a periodic spatial structure. In such cases, the Fermi surface
volumes are well-defined modulo the Brillouin zone volume
$\Omega_B$; as a consequence all of our statements on the
constraints on Fermi surface volumes are modulo $\Omega_B$. The
situation is identical to electrons moving in a periodic potential
considered by Luttinger originally \cite{luttinger}. We note in
passing that the possible realization of the FFLO state in
CeCoIn$_5$ \cite{radovan,bianchi} is currently being investigated
experimentally.

\section{One dimensional systems}
\label{sec:1d}

We now turn our discussion to one-dimensional (1D) systems, where
there are no well-defined quasiparticles, but there can still be
well-define Fermi wavevectors and corresponding Luttinger's theorem
\cite{yoa}. In the following we consider a 1D lattice version of
$H$:
\begin{eqnarray}
H_{1D}&=&-t_f\sum_{i,\sigma}(f^\dagger_{\sigma i}f_{\sigma
i+1}+h.c.)
-t_b\sum_i(b^\dagger_{i}b_{i+1}+h.c.)\nonumber\\
&-&g\sum_{i}(f^\dagger_{\uparrow i}f^\dagger_{\downarrow
i}b_i+h.c.)+H_{int}, \label{h1d}
\end{eqnarray}
where $i$ is the site index and $H_{int}$ involves fermion or boson
number operators only. Such models have been considered recently
\cite{sheehy,citro} in the context of mixture and coherence between
fermionic atoms and bosonic molecules in 1D. Here we consider the
most generic case that $N_\uparrow-N_\downarrow\ne 0$ and the
particle fillings are incommensurate with the lattice, so that
neither spin nor charge gap is possible; generalizations to
commensurate cases are straightforward. In the absence of the
bosonic degrees of freedom, Eq.~(\ref{h1d}) describes a Luttinger
liquid with two (charge and spin) linearly dispersing gapless modes,
and there are two Fermi wavevectors associate with $\uparrow$ and
$\downarrow$ fermions, whose magnitudes are determined individually
by $N_\uparrow$ and $N_\uparrow$, which are separately
conserved\cite{yoa}. In the presence of bosonic degrees of freedom
and Bose-Fermi mixture as described by $H_{1D}$ in Eq.~(\ref{h1d}),
it was pointed out recently \cite{sheehy} that a new gapless mode
may appear due to the condensation (or formation of an additional
Luttinger liquid) of the bosons. This ``decoupled spin-gapless
phase" \cite{sheehy} is the 1D analog of the high-D phases with a
Bose condensate discussed earlier. We show below that there exist 1D
versions of Luttinger's theorem that completely determine the Fermi
wavevectors in the absence of this new gapless mode, while in the
presence of it (or in the decoupled spin-gapless phase) the
Luttinger's theorem only gives a constraint on the Fermi wavevectors
in a manner analogous to the high D cases discussed earlier.

Following Ref.~\onlinecite{yoa}, we consider a system with linear
size $L$ and impose periodic boundary condition. We introduce twist
operators that are appropriate to $H_{1D}$ in Eq. (\ref{h1d}), in
spin and charge sectors respectively:
\begin{eqnarray}
U_s&=&\exp\left[i{2\pi\over L}\sum_j
j(n_{j\uparrow}-n_{j\downarrow})\right],\\
U_c&=&\exp\left[i{2\pi\over L}\sum_j
j(n_{j\uparrow}+n_{j\downarrow}+2n_{jb})\right].
\end{eqnarray}
It is straightforward to show that $U_{s,c}|0\rangle$ generates
states with momenta
\begin{eqnarray}
k_s&=&{2\pi\over L}(N_\uparrow-N_\downarrow)={2\pi\over L}N_s,\label{spin}\\
k_c&=&{2\pi\over L}(N_\uparrow+N_\downarrow+2N_B)={2\pi\over
L}N_c,\label{charge}
\end{eqnarray}
whose energies vanish as $1/L$ in the limit $L\rightarrow\infty$.
These gapless excitations at finite wavevectors correspond to
combinations of $2k_F$ excitations in Luttinger liquids \cite{yoa}.

In the absence of the new gapless mode due to the bosons, there are
one spin mode and one charge mode that are gapless, and the two
Fermi wavevectors can be uniquely determined by (\ref{spin}) and
(\ref{charge}):
\begin{equation}
2k_{F\sigma}=(k_c+\sigma k_s)/2=(\pi/L)(N_c+\sigma N_s),
\end{equation}
where $\sigma=\pm 1$ for $\uparrow/\downarrow$. On the other hand
when there is a new gapless mode due to the bosons, there are one
spin mode and {\em two} charge modes that are gapless; in this case
there is an additional Fermi wavevector associated with this new
charge mode (which may be interpreted as the Fermi wave vector of
the condensed bosons). In this case Eqs.~(\ref{spin}) and
(\ref{charge}) can no longer determine all the Fermi wavevectors
uniquely. However Eq. (\ref{spin}) still poses a constraint for the
Fermi wavevectors that correspond to the fermions:
\begin{equation}
2(k_{F\uparrow}-k_{F\downarrow})=k_s=(2\pi/L)N_s.
\end{equation}
These results are completely analogous to those obtained for high D
systems.

It has been shown \cite{polkovnikov} recently that correlations in
trapped 1D cold atom systems can be probed through interference
experiments. In particular, it was found that the intensity of
interference patterns can be used to directly measure the square of
the equal time single particle Green's function \cite{polkovnikov}.
For fermions, the Green's function is oscillatory with
characteristic wave vector $k_F$. As a result $k_F$ can be extracted
from such experiments, and our results above on 1D systems with
Bose-Fermi mixture can be tested directly.

\section{Quantum phase transitions}
\label{sec:qpt}

We conclude this paper by presenting the field theories controlling
the various quantum phase transitions and multicritical points in
Fig.~\ref{phase} in the following subsections. Some aspects of these
theories were also discussed in Ref.~\onlinecite{powell}.

\subsubsection{2FS/1FS}

We begin with the transition between the 2FS and 1FS phases at large
$h$. Here a Fermi surface gradually shrinks to zero volume, in the
absence of any bosonic condensate. Denoting the fermionic
quasiparticle excitations of this small Fermi surface by $\psi$, and
following the analysis of the dilute Fermi gas critical point in
Chapter 11 of Ref.~\onlinecite{book}, we initially guess that the
critical theory is given by the free fermion form
\begin{equation}
S_\psi = \int d^d x d \tau \, \psi^\dagger
\left(\frac{\partial}{\partial \tau} - \frac{1}{2m_\psi} \nabla^2 +
s \right) \psi,
\end{equation}
where $\tau$ is imaginary time, and $s$ is the parameter which tunes
across the transition. In this case, the quantum critical point is
at $s=0$, and the small Fermi surface is present for $s>0$. The
simplest four Fermi interaction must involve two gradients because
of the Pauli principle, and is of the form $\sim (\psi^\dagger
\nabla \psi )^2$. A scaling analysis with dynamic exponent $z=2$
\cite{book} shows that the dimension of this term is $-d$, and so
such interactions are always irrelevant. However, here we have a
second large Fermi surface present, and we have to allow for
long-range interactions induced by the fluctuations of this Fermi
surface. The oscillations of this second Fermi surface will couple
to $\psi^\dagger \psi$, and integrating these out following Hertz
\cite{hertz} we obtain the following long-range interaction (for
$d>1$)
\begin{equation}
\lambda \int_{{\bf k}, \omega} [ \psi^\dagger \psi]_{{\bf k},
\omega} \, \frac{|\omega|}{k} \, [ \psi^\dagger \psi]_{-{\bf k},
-\omega} \label{zs}
\end{equation}
Power-counting with $z=2$ shows that $\lambda$ has scaling dimension
$1-d$, and so is irrelevant. The $d=1$ case requires special
treatment, and we will not consider it here.

\subsubsection{2FS+BEC/1FS+BEC}

This is just like the 2FS/1FS transition above, except that a
background Bose condensate is present. Fluctuations in the superflow
of this condensate induce long-range interactions, which could be
important at the critical point. With $\varphi$ being the phase of
the condensate, the coupling to the $\psi$ excitations of the
current-current type \cite{son}: $\partial_i \varphi (\psi^\dagger
\partial_i \psi - \partial_i \psi^\dagger \psi)$. Integrating out
the $\varphi$, we now generate the interaction
\begin{eqnarray}
&& \int_{{\bf k}_1 , {\bf k}_2 , {\bf q}, \epsilon_1, \epsilon_2,
\omega} \psi^\dagger ({\bf k}_1 + {\bf q}, \epsilon_1 + \omega)
\psi ({\bf k}_1, \epsilon_1) \nonumber \\
&&\times\frac{({\bf k}_1 \cdot {\bf q}) ({\bf k}_2 \cdot {\bf
q})}{q^2} \psi^\dagger ({\bf k}_2 - {\bf q}, \epsilon_2 - \omega)
\psi ({\bf k}_2, \epsilon_2). \label{superflow}
\end{eqnarray}
Again, simple power-counting shows that this term has scaling
dimension $-d$, and so is irrelevant. So the critical theory remains
the free fermion theory $S_\psi$.

\subsubsection{BEC/1FS+BEC}

The considerations for this are identical to the 2FS+BEC/1FS+BEC
transition above, as both involve the disappearance of a Fermi
surface in the presence of bosonic condensate.

\subsubsection{BEC/2FS+BEC}

This is a novel transition involving the simultaneous appearance of
two Fermi surfaces at a spherical minimum in the fermionic
quasiparticle dispersion. This was considered in a separate paper
\cite{ys} for the case of non-singular (or short-range) fermionic
interactions. The superflow fluctuations will also introduce here an
interaction between the fermions of the form in
Eq.~(\ref{superflow}). Because we are now interested in fermions at
large momentum Fermi surface, the ${\bf k}_{1,2}$ in
Eq.~(\ref{superflow}) will be replaced by $k_0 {\bf n}$, where $k_0$
is the minimum of the fermion dispersion (that forms a sphere; it
also becomes the Fermi wavevector once fermions start to populate
these modes), and ${\bf n}$ is its normal direction. This changes
the power-counting in the scaling dimension of this coupling. We
discuss the renormalization group (RG) flow of such anisotropic,
long-range interactions below.

As discussed in Ref. \onlinecite{ys}, the relevant fermionic modes
near the critical point are those within a thin spherical shell in
momentum space: $k_0 -\Lambda < |{\bf k}| < k_0 +\Lambda$, where
$\Lambda\ll k_0$ is a cutoff scale. Phase space restriction thus
divide the possible scattering processes into two
classes, which need to be treated separately: \\
({\em i\/}) The Cooper channel, in which ${\bf k}_1\approx -{\bf
k}_2$, and the momentum transfer $q$ can be as large as $O(k_0)$. In
this channel the interaction of Eq. (\ref{superflow}) simply reduces
to those studied in Ref. \onlinecite{ys}, and the results there can
be applied directly to the present case.\\
({\em ii\/}) Forward scattering channel, in which $|{\bf k}_1+{\bf
k}_2|\gg \Lambda$, and the magnitude of momentum transfer is
constrained to be $|{\bf q}|\lesssim \Lambda$. In the absence of
long-range interactions, we can then take the limit $|{\bf q}|
\rightarrow 0$, and assume that there is no further dependence on
its orientation ${\bf q}/|{\bf q}|$. However, with the anisotropic
long-range interactions induced by superflow fluctuations in
Eq.~(\ref{superflow}), the limit of vanishing $|{\bf q}|$ can be
taken, but there remains a residual dependence on ${\bf q}/|{\bf
q}|$. Here we perform an RG analysis of the most general interaction
of fermions on the Fermi surface dependent upon ${\bf q}/|{\bf q}|$,
and for simplicity focus on two-dimensional (2D) systems; the
situation in 3D is expected to be similar qualitatively. Such an
interaction takes the following form in 2D:
\begin{equation}
V({\bf k}_1, {\bf k}_2, {\bf q})=\sum_nV_n(\theta)\cos(n\phi_{\bf
q}), \label{forward}
\end{equation}
where $\theta$ is the angle between ${\bf k}_1$ and ${\bf k}_2$, and
$\phi_{\bf q}$ is the angle between ${\bf q}$ and ${\bf k}_1+{\bf
k}_2$. For short-range interactions only the $n=0$ term exists, and
this case was analyzed in Ref.~\onlinecite{ys}. The interaction in
Eq. (\ref{superflow}) generates a term with $n=2$. We now perform a
one-loop RG analysis \cite{ys} of the interactions in
Eq.~(\ref{forward}), and obtain the following flow equation:
\begin{equation}
{dV_n(\theta)\over d\log s}=\delta_{n,0}\beta_0[V],
 \label{flow}
\end{equation}
where $s$ is the cutoff rescaling factor and
\begin{equation}
\beta_0[V]={-2m^*\over
\pi^2|\sin\theta|}\int_{-1}^1{|\sum_nV_n(\theta)\cos[n\phi(t)]|^2\over
1+t^2}dt, \label{flow2}
\end{equation}
in which $\phi(t)=\arctan[(t-\cos\theta)/\sin\theta]-\theta/2$, and
$m^*$ is a parameter that parameterize the fermion dispersion near
the minimum: $\epsilon(k)=(k-k_0)^2 / (2m^* )$. It is clear from
Eq.~(\ref{flow}) that only $V_0$ gets renormalized under RG
transformations, while interactions with $n\ne 0$ remains marginal.
Physically this is because the renomalization is due to multiple
virtual scattering processes whose momentum transfers are of order
the cutoff $\Lambda$, resulting in renomalization that is not
sensitive to the net momentum transfer much smaller than $\Lambda$
at low-energy. Thus the $n \neq 0$ quasiparticle interactions remain
at their bare values, and are finite and non-universal in the
low-energy limit. For the case of the quantum phase transition of
interest in this subsection, the $n=2$ interaction in
Eq.~(\ref{superflow}) therefore acquires no loop corrections.
Furthermore, in the presence of $V_{n\ne 0}$, $V_0$ will always be
driven negative and flow away, even if it is initially positive (or
repulsive). The situation was very different when only the
short-range interaction $V_0$ was present; in that case it was found
\cite{ys} that for repulsive interactions, it flows to zero
logarithmically, which leads to an effective quasiparticle
interaction that takes a universal form in the low-energy limit due
to this renormalization. Here, we observe from Eq.~(\ref{flow2})
that the $n=2$ interactions leads to a {\em run-away} flow of $V_0$
that has no fixed point. Physically, this suggests that either the
transition from BEC to 2FS+BEC is first order, or that the
instability of the BEC in this case is toward a state with other
symmetry properties, like the FFLO state. We note that related
observations were made in Ref. \onlinecite{son} based on quite
different considerations.

\subsubsection{1FS/1FS+BEC}

This involves the appearance of a BEC in the presence of Fermi
surface. The BEC onset is described by the dilute Bose gas theory
discussed in Chapter 11 of Ref.~\onlinecite{book}, and in
Ref.~\onlinecite{dunkel}. The action is
\begin{eqnarray}
S_b &=& \int d^d x \int d \tau \Biggl[ b^\dagger \frac{\partial
b}{\partial \tau} - \frac{1}{2m_b} b^\dagger \nabla^2 b + s |b|^2
\nonumber \\ &~&~~~~~~~~~~~~~~~~~~~~~~~~~+ \frac{u}{2} |b|^4
\Biggr], \label{scb}
\end{eqnarray}
where $s$ is again the tuning parameter, and the quantum critical
point is at $s=0$. The scaling dimension of the quartic coupling $u$
is $(2-d)$, and so is formally irrelevant for $d>2$. However, this
coupling is {\em dangerously\/} irrelevant, because a $u>0$ is
required to stabilize the system for $s<0$. Couplings to Fermi
surface fluctuations will induce a long-range interaction analogous
to Eq.~(\ref{zs}) among the bosons
\begin{equation}
\lambda \int_{{\bf k}, \omega} [|b|^2]_{{\bf k}, \omega} \,
\frac{|\omega|}{k} \, [ |b|^2 ]_{-{\bf k}, -\omega}. \label{zsb}
\end{equation}
Just as in Eq.~(\ref{zs}) however, power-counting near a $z=2$
transition shows that this coupling has scaling dimension $1-d$, and
so is irrelevant.

\subsubsection{2FS/2FS+BEC}

The considerations for this are nearly identical to the 1FS/1FS+BEC
transition discussed above. Here, we also have to consider the decay
of a boson into two fermions, one each on the respective Fermi
surfaces. However, this is a low energy process only at finite
wavevectors which connect the the two Fermi surfaces. Consequently
it can be safely neglected for the low momentum $b$ critical theory.

\subsubsection{Multicritical point M}

The theory for this point is essentially the direct sum of the
theories for the two transition lines which intersect at M. There is
a bosonic critical mode, $b$, described by $S_b$, and a fermionic
critical mode $\psi$, described by $S_\psi$. These two critical
modes can interact with each other via the contact term
\begin{equation}
g \int d^d x d\tau |b|^2 \psi^\dagger \psi,
\end{equation}
and $g$ has dimension $(2-d)$. So it will have appreciable effects
in $d=2$ which can be computed along the lines of
Ref.~\onlinecite{dunkel}. The coupling $g$ will also be very
important in $d=1$.

\subsubsection{Multicritical point L}

This is a `Lifshitz' point, which was considered in
Ref.~\onlinecite{son}, albeit with a different perspective. This is
the point where the fermionic dispersion minimum moves from ${\bf k}
= 0$ to a non-zero $k$. Consequently, the dispersion at small ${\bf
k}$ is $\sim k^4$, and the multicritical theory has $z=4$:
\begin{equation}
S_{\psi,L} = \int d^d x d \tau \, \psi^\dagger
\left(\frac{\partial}{\partial \tau} +  \nabla^4  \right) \psi.
\end{equation}
The simplest four Fermi interaction is again $\sim (\psi^\dagger
\nabla \psi )^2$, and a scaling analysis with $z=4$ shows that this
has scaling dimension $(2-d)$. The superflow fluctuations will also
induce a long range interaction as in Eq.~(\ref{superflow}), and
this again has scaling dimension $(2-d)$. So these interactions are
irrelevant for $d>2$, while a detailed analysis of logarithmic
corrections is necessary in $d=2$. We do not present this here, but
it can be carried out as in Ref.~\onlinecite{dunkel}.

\acknowledgements

This work was supported by National Science Foundation grants
DMR-0537077 (SS), DMR-0225698 (KY), and a Florida State University
Research Foundation Grant (KY).

\end{document}